\documentstyle[preprint,aps,tighten,epsf,floats]{revtex}
\newcommand{\ben}{\begin{displaymath}}
\newcommand{\een}{\end{displaymath}}
\newcommand{\be}{\begin{equation}}
\newcommand{\ee}{\end{equation}}
\newcommand{\bea}{\begin{eqnarray}}
\newcommand{\eea}{\end{eqnarray}}

\newcommand{\fign}[1]{\label{#1}}

\begin{document}
\draft
\title{
Is Heavy Baryon Approach Necessary?}
\author{ J. Gegelia${ }^a$,
G. Japaridze and X.Q.  Wang${ }^b$ }
\address{${ }^a$ School of Physical Sciences, Flinders University of South 
Australia, \\ Bedford Park, S.A. 5042, Australia. 
\\ 
${ }^b$ Department of Physics, Center for Theoretical studies of Physical 
Systems, \\ Clark 
Atlanta University, Atlanta, GA 30314} 
\date{\today}
\maketitle 

\begin{abstract}
It is demonstrated  that using an
appropriately chosen renormalization condition one can respect power counting
within the relativistic baryon chiral perturbation theory  without appealing to
the technique of the heavy baryon 
approach. Explicit calculations are performed for
diagrams including two-loops. It is argued that the introduction of the  heavy
baryon chiral perturbation theory was useful but not necessary.
\end{abstract}
\medskip
\medskip
PACS number(s):
03.70.+k
11.10.Gh,  
12.39.Fe,  
\section{introduction}

Chiral symmetry is of fundamental importance in the low energy dynamics of
strongly 
interacting particles. 
Using this symmetry Weinberg and Gasser and 
Leutwyler have developed 
Chiral Perturbation Theory, a systematic and feasible scheme for calculating
processes of meson-meson interactions \cite{weinberg,gasser1}.
The Chiral Perturbation Theory possesses the feature of consistent power
counting which allows systematic perturbative calculations.

The nontrivial problem appeared after Gasser, Sainio and \v Svarc 
considered processes with a single baryon \cite{gasser}. 
They noticed that there is no 
consistent power counting when a baryon is included; higher order loops
contribute to low order (in small expansion parameters)
calculations. Performing 
calculations at any given order of the chiral expansion 
one needs to include contributions of diagrams with an increasing
(up to infinity) number of loops. 

To avoid this drawback which makes the results of perturbative calculations
unreliable, Jenkins and Manohar suggested to consider an extremely
nonrelativistic limit of  
the original relativistic field-theoretical model \cite{jenkins}. Integrating
out heavy degrees of freedom and expanding the resulting effective action in
inverse powers of the 
baryon mass $M$, they developed Heavy Baryon Chiral Perturbation Theory
(HB$\chi$PT). 
In the framework of HB$\chi$PT the power counting is restored and thus the
problem  
of the relativistic treatment of the sector with baryon number $1$,
encountered in  
\cite{gasser} is circumvented. The revival of power counting is
traded for 
explicit relativistic invariance. Nowadays HB$\chi$PT is an 
effective method of 
calculation of different processes involving electro-magnetic and strong
interactions (for a review and references see \cite{bernard} and
\cite{fettes}). 

In this paper we investigate whether the violation of power counting is
an intrinsic feature of the relativistic effective theory of pion-nucleon
interactions or is only an
artifact of the particular method of calculation.

In this connection let us remind the reader that the problem in the relativistic
approach of multi-loop diagrams 
contributing to low order calculations \cite{gasser} was actually encountered
in the $\overline{MS}$ scheme. This scheme puts the
effective cut-off equal to the largest involved mass scale, i.e. the nucleon
mass and 
violates the power counting.

On the other hand, for processes involving an arbitrary number of nucleons
Weinberg suggested the usage of renormalization points of the order of 
external momenta or less \cite{weinberg1,weinberg2}. Such schemes put the
effective cutoff for loop integrals of the order of external momenta and make
power counting applicable for loop diagrams. While Weinberg considered a
non-relativistic effective field theory, the same idea of an appropriate choice
of
renormalization condition can be useful in relativistic theory as well. 
As was discussed in \cite{tang1,tang2,gj}, parts of relativistic diagrams
responsible for the violation of power counting can be altered by adding
counter-terms. Hence they can be removed by choosing an appropriate
renormalization condition.

In the present paper we reexamine the question of validity of chiral power
counting in 
relativistic baryon chiral perturbation theory. We work in exact chiral limit
and suppress the isospin. 
The resulting expansion in small momenta simulates  
essential features of chiral perturbation theory for pion-nucleon 
interaction (it is straightforward to  show that the isospin and the small
non-vanishing mass  
of the pseudoscalar particle just complicate calculations and do not affect
our results).  
Calculating one and two-loop diagrams we demonstrate 
that by choosing an appropriate subtraction scheme  
one can respect power counting in the relativistic theory without
appealing to the heavy baryon technique. In other words, within the suggested
scheme power counting does not fail when baryons are introduced. 

In ref. \cite{tang1,tang2,becher,lutz} it has been argued  that consistent power
counting can exist within the relativistic scheme. 
Our approach is substantially different,
we consider a conventional renormalization technique \cite{collins} and in
contrast with 
\cite{tang1,tang2,becher} we do not split loop integrals
into soft and hard parts and also we do not need to include an infinite 
number of counter-terms while performing renormalization at any finite 
order.

\section{One-loop approximation}

We consider a field theoretical model described by the Lagrangian:

\begin{equation}
{\cal L}=-{1\over 2}\phi\partial_{\mu}\partial^{\mu}\phi 
+\bar\psi\left( i\gamma^\mu\partial_\mu-M\right)\psi 
-g\bar\psi\gamma_5\gamma^\mu\psi\partial_\mu\phi +\bigtriangleup {\cal L}
\label{lagrangian}
\end{equation}
In eq. (\ref{lagrangian}) $\psi$ is a fermion field with mass $M$,  $\phi$ is a
neutral massless pseudoscalar
field, 
$g$ is a coupling constant and $\bigtriangleup {\cal L}$ includess all 
counter-terms necessary to remove 
divergences. 
We use dimensional regularization with $n$ being the space-time dimension.

Lagrangian (\ref{lagrangian}) suggests that analogously to the meson chiral
perturbation theory \cite{weinberg} for diagrams containing one fermion
line we can have a (naive?) power counting. We assign +4 powers (of small
momenta) 
to each loop integration, +1 power to each derivative occurring in the
interaction term, -2 powers to each scalar propagator and -1 power
to each fermion propagator. Thus a resulting power 
$N_{i}$ is assigned to each particular diagram $i$. We will say that diagram $i$
obeys power counting if the leading term  
of the result of actual calculation depends on a small momentum $k$ as 
$k^{N_{i}}f(k)$, where $f(k)$ is a constant or logarithmic function of $k$. 

This power counting is badly 
violated if the $\overline {MS}$ scheme is used; higher order loops
do not lead to higher orders in $k$.  

Below we demonstrate that
the breakdown of power counting is the result of using the $\overline{MS}$
scheme, and, by applying an appropriately chosen renormalization condition 
it is possible to retain the power counting within relativistic 
theory. 

\begin{figure}[t]
\hspace*{3.5 cm}  \epsfxsize=10cm\epsfbox{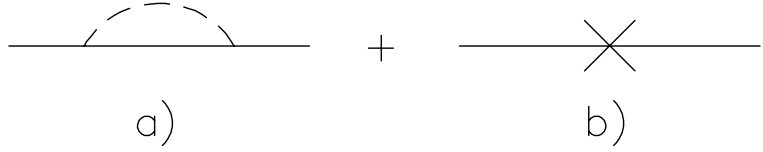}
\caption{\fign{se}{\it (a) One-loop correction to the fermion propagator
and (b) corresponding counter-term diagram. Solid line corresponds to fermion
and dashed line to pseudoscalar}}
\end{figure}

Let us consider the one-loop correction to the fermion propagator depicted in
Fig.1 (a). The corresponding expression is: 
\be
\Sigma_1=-g^2\int {d^nq \over (2\pi)^n} {\gamma_5\not q\left( \not p+\not q
+M\right)\gamma_5\not q\over \left[ q^2+i\epsilon\right]\left[ 
(p+q)^2-M^2+i\epsilon\right]},
\label{int1}
\ee
where $\not p=p^\mu \gamma_\mu$, $p^\mu =Mv^\mu +k^\mu$ is off mass-shell
momenta of the fermion, 
$v_\mu v^\mu =1$ and $k^\mu (<<M)$ is a small quantity. 
Eq. (\ref{int1}) can be reduced to the following form: 
\be
\Sigma_1=g^2 MJ(01)-g^2 (n-1)J^{n+2}(11)\not p
\label{int1m}
\ee
where $J(01)$ and $J^{n+2}(11)$ are given 
in the Appendix.

Substituting the values of $J(01)$ and $J^{n+2}(11)$  into
eq. (\ref{int1m}) we obtain:
$$
\Sigma_1=-M{ig^2\over (4\pi)^{n/2}}\left( M^2\right)^{n/2-1}
\Gamma\left( 1-n/2\right)-\not p{3\over 2}{ig^2\over (4\pi)^{n/2}}
\left( M^2\right)^{n/2-1}
\Gamma\left( 1-n/2\right)
$$
$$
-\not pp^2{ig^2\over (4\pi)^{n/2}}\left( M^2\right)^{n/2-2}
{\Gamma\left( 2-n/2\right)\over n-2}
-\not p{ig^2\over (4\pi)^{n/2}}\left( M^2\right)^{n/2-1}
{\Gamma\left( 3-n/2\right)\over (2-n)(3-n)}\left( 1-z\right)^2
$$
$$
-\not p{ig^2\over (4\pi)^{n/2}}\left( M^2\right)^{n/2-1}
{\Gamma\left( 4-n/2\right)\over (2-n)(3-n)(4-n)}\left( 1-z\right)^3
\times { }_2F_1\left( 1,4-{n/2};5-n;1-z\right)
$$
\be
+\not p{ig^2\over (4\pi)^{n/2}}\left( M^2\right)^{n/2-1}
\Gamma\left( n/2\right)\Gamma\left( 2-n\right)
\left( 1-z\right)^{n-1}
\times { }_2F_1\left( {n/2},n;n;1-z\right)
\label{int1c}
\ee
In eq. (\ref{int1c}) ${ }_2F_1$ is the Gauss hypergeometric function
\cite{brichkov}  
and we have introduced dimensionless quantity $z\equiv p^2/M^2$.

In order to carry out the renormalization procedure it is necessary to add 
to eq. (\ref{int1c}) contributions from the
counter-terms. As was mentioned above all the necessary
counter-terms are 
included in the Lagrangian and the corresponding contribution reads:
\be
\delta\Sigma_1=g^2 \delta_1 +g^2\delta_2\not p+g^2\delta_3p^2\not p
+g^2\delta_4p^4\not p
\label{cct1}
\ee
Equation (\ref{int1c}) fixes only the divergent parts of $\delta_{i}$ which
have to be the same for every scheme and the choice of the particular
renormalization scheme corresponds to the choice of finite parts of
counter-terms. 
Note that $\delta_4$ is finite. We are free to add such finite counter-terms
exploiting our freedom of the choice of the renormalization scheme. 

We choose $\delta_4$ and the finite parts of $\delta_1$, $\delta_2$ and 
$\delta_3$ so as to cancel the first four
terms in eq. (\ref{int1c}). The remaining expression admits the limit 
$n\rightarrow 4$ and the renormalized self energy is: 
\be
\Sigma_1^R=-{ig^2\over 32 \pi^2}M^2\not p\frac{(1-z)^{3}}{z}\left[ \frac{1}{z}
\ln (1-z)+1 \right] 
\label{int1cR}
\ee 
Power counting states that $\Sigma_1^R$ is expected to be of a third order in
$k$, and since $1-z=O(k)$, it 
follows that eq. (\ref{int1cR}) is in agreement with this prediction. 

On the other hand it is straightforward to show that applying 
$\overline {MS}$ to eq. (\ref{int1c}) and taking into
account the mass and wave function renormalizations we get 
$\Sigma^R_{1(\overline {MS})}\sim k$.

\section{Two-loop analysis}

\begin{figure}[t]
\hspace*{1.5cm}  \epsfxsize=14cm\epsfbox{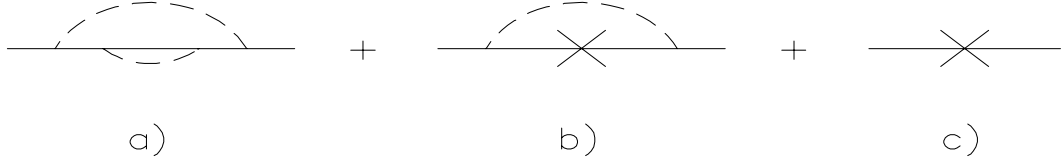}
\caption{\fign{se}{\it (a) Two-loop correction to the fermion propagator, (b)
corresponding diagram with one loop sub-diagram replaced with $g^2$-order
counter-terms and 
(c) $g^4$-order counter-term diagram.}}
\end{figure}

Two-loop corrections to the fermion propagator  have more complicated 
structures.  
For illustrating purposes it is enough to consider a diagram depicted in
Fig.2 (a). The corresponding expression is:
\be
\Sigma_{2(a)}=ig^4\int {d^nq_1d^nq_2 \over (2\pi)^{2n}} {\gamma_5\not q_1
S_F\left( p+q_1\right)\gamma_5\not q_2S_F\left( p+q_1+q_2\right)
\gamma_5\not q_2S_F\left( p+q_1\right)\gamma_5\not q_1\over 
\left[ q_1^2+i\epsilon\right]\left[ q_2^2+i\epsilon\right]}
\label{int2}
\ee
Simplifying eq. (\ref{int2}) we obtain:
$$
\Sigma_{2(a)}=ig^4\Biggl \{ \left( M^2-p^2\right) \gamma^{\mu }J_{\mu}(1101) 
-2\not p\left[ {M^2-p^2\over 2n}J(1101)-{2\over n}p_{\mu}J^{\mu }(1101)
+\left( 1-{1\over n}\right)p^2C_2\right]
$$
$$
+2MJ(01)J(01)-4M^2\gamma^{\mu }J_{\mu}(11)J(01)
+4M^2(\not p+M)J(01)J(02)
$$
\be
+4M^2\left( p^2-M^2\right)\gamma^{\mu }J_{\mu}(12)J(01)\Biggr \},
\label{sigma2}
\ee
where $J_{\mu}(1101)$, $J(1101)$, $C_2$, $J_{\mu}(11)$, $J(02)$ and
$J_{\mu}(12)$ are given in the Appendix.

The renormalization procedure requires that we add to eq. (\ref{sigma2})
expressions corresponding 
to Fig.2 (b) and Fig.2 (c). The expression for Fig.2 (b) is readily obtained
replacing the one-loop 
subdiagram of Fig.2 (a) by counter-terms of order $g^{2}$; diagram Fig.2 (c)
corresponds 
to the contribution of counter-terms of order $g^{4}$. 
Calculating diagram Fig.2 (b) and adding to expression (\ref{sigma2}) we
obtain: 

\begin{equation}
\Sigma_{2(a)}+\Sigma_{2(b)}=iM^{2}g^4\Biggl \{ \left( 1-z\right) \gamma^{\mu }J_{\mu}(1101) 
-2\not p\left[ {1-z\over 2n}J(1101)-{2\over nM^{2}}p_{\mu}J^{\mu }(1101)
+\left( 1-{1\over n}\right)zC_2\right]\Biggr \}
\label{sigmatilde}
\ee

We express the counter-terms corresponding to Fig.2 (c) as
\be
\delta\Sigma_2=g^4\not p \Biggl( \delta_5 p+p^2\delta_6 p+p^4\delta_7
+p^6\delta_8+p^8\delta_9 \Biggr)
\label{ct2}
\ee
As in the case of the one-loop correction, the divergent parts of $\delta_{i}$
are uniquely fixed from the requirement that renormalized self energy is free
of  divergences. 
We have introduced finite terms $\delta_8$ and $\delta_9$. 
These terms, together with the finite parts of $\delta_5$, $\delta_6$ and $\delta_7$
are fixed below by the renormalization condition. 

Next we express $\delta\Sigma_2$ in terms of $z$ and $M^2$:
$$
\delta\Sigma_2=\not pM^2g^4\Biggl \{ \delta_5/M^2+\delta_6+M^2\delta_7
+M^4\delta_8+M^6\delta_9-(1-z)\left[ \delta_6+2M^2\delta_7
+3M^4\delta_8+4M^6\delta_9\right]
$$
\be
+(1-z)^2\left[ M^2\delta_7+3M^4\delta_8+6M^6\delta_9 \right]-(1-z)^3\left
[ M^4\delta_8+4M^6\delta_9\right]+(1-z)^4M^6\delta_9\Biggr \}, 
\label{tlctc}
\ee
and we expand the analytic part of eq. (\ref{sigmatilde}) in $(1-z)$ and add 
$\Sigma_{2(a)}+\Sigma_{2(b)}$ to $\delta\Sigma_2$.
We define the renormalization scheme via the condition that 
$\delta_8$, $\delta_9$ and finite parts of $\delta_5$,
$\delta_6$, $\delta_7$ are fixed so as to exactly cancel the first five terms
(up to and including $(1-z)^4$) in the expansion of the analytic
part. Performing all these calculations we get:
$$
\Sigma_2^R=ig^4\not p {\left( M^2\right)^{n-2}\over (4\pi)^n}
\Biggl \{ -{3\over 4}\left[ {\rm Li}_2(1-z)-(1-z)-{(1-z)^2\over 4}\right]
+{z(1-z)^3\over 12}-{(1-z)^4\over 48z}+{29\over 192}(1-z)^4
$$
\be
+{(1-z)^5\over 12z}+\ln (1-z) \Biggl [ -{9\over 8}(1-z)^2
-{3\over 4}\ln z-{3\over 4}z(1-z)-{(1-z)^3\over 4z}+{(1-z)^4\over 16z^2}
\Biggr ]\Biggr \}
\label{sigmaren}
\ee
In eq. (\ref{sigmaren}) ${\rm Li}_2$ is the dilogarithm function
\cite{abramowitz}.

It is straightforward to verify that the coefficient function 
of $\ln (1-z)$ as well as the analytic part of $\Sigma_2^R$ are of order
$(1-z)^{5}$.  Therefore the two-loop correction to fermion self energy 
satisfies power counting.

On the other hand, applying $\overline {MS}$ scheme to $\Sigma_{2(a)}$ and
taking 
into account the mass and wave function renormalizations we obtain 
$\Sigma_{2(\overline {MS})}^R\sim k$.
Therefore, as was already observed in \cite{gasser}, we see that in the
$\overline {MS}$ 
scheme two-loop and one-loop contributions in the fermion self energy are 
of the same order in $k$. 

Let us summarise our approach to relativistic baryon chiral
perturbation theory.

To remove divergences from Feynman diagrams we use the forest
formula of 
Zimmermann \cite{collins}. The forest formula is applied to individual
diagrams 
and substracts the overall divergence as well as 
the divergences corresponding  to all subdiagrams.
These subtractions can be implemented as counter-terms in the Lagrangian
\cite{collins}.   
In performing 
actual calculations we do not necessarily need the explicit expressions for
counter-terms;  
the substraction scheme can be specified by pointing out the prescription for 
the finite parts. In this scheme the parameters of the Lagrangian are considered
as finite renormalized coupling constants.
In relativistic chiral perturbation theory instead of the widely used
$\overline{MS}$ scheme we should apply a substraction scheme 
which respects power counting.
To do so we suggest the following strategy. First renormalise one-loop diagrams
by expanding the analytic (in small momenta) parts in powers of a small momentum
and perform covariant
subtraction so as to cancel first few terms in the above mentioned expansion
and respect the power counting. According to \cite{gj} the non-analytic parts 
readily obey power counting. For two-loop diagrams we first subtract one-loop
subdiagrams and after expand analytic parts in powers of small momenta and
again perform covariant subtraction so as to cancel the first few terms in the
above expansion of the analytic part and
respect the power counting. The non analytic parts 
remaining after the subdiagrams are subtracted respect power counting. 
For the three-loop diagrams the strategy is the same: first substract one and
two-loop subdiagrams, expand the analytic parts, etc. This iterative procedure
is well defined for any number of loops.      
Within the suggested subtraction scheme the higher order diagrams do not
contribute to lower order calculations. Consequently, coupling constants 
defined via low order calculations are not affected by higher order
corrections.  
Our results remain valid when the pseudoscalar
particle acquires a small mass. 
Note that in realistic model of baryon chiral perturbation theory one should fix
finite parts of different counter-terms (specify prescriptions for subtractions)
so as to respect corresponding Ward identities. As far as there are no
anomalies it is always possible to satisfy this requirement within our scheme
which is nothing else than the conventional renormalization with particular
renormalization condition.   

\section{Conclusion}

We have demonstrated that by choosing an adequate renormalization scheme it is 
possible to retain the power counting in relativistic baryon chiral perturbation
theory. 
Hence there is no necessity to invoke the heavy baryon approach. 
Although HB$\chi$PT substantially 
simplifies  the calculations for many physical quantities, the corresponding
perturbation series fails to converge in part of the low energy region
\cite{bernard} (this problem has 
been resolved by Becher and Leutwyler using the ``infrared
regularization'' technique \cite{becher}). The original
relativistic approach never encounters this problem.   
Hence both approaches enjoy their advantages and each has a full right to exist.

\medskip
\medskip
\newpage

{\bf ACKNOWLEDGEMENTS} 

This work was carried out while one of the authors (J.G.) was supported by
Flinders University Research Scholarship
at Flinders University of South Australia.

This work was supported in part by 
Air Force Office of Scientific Research under
Grant No. F4962-96-1-0211 and Army Research Office under Grant No.
DAAH04-95-1-0651. One of us (G. J.) is grateful to A. Msezane and C. Handy 
for their support.

\appendix
\section{}
$$
z\equiv {p^2\over M^2}
$$

\be
J(01)={1\over (2\pi)^n}\int {d^nq\over \left[ q^2-M^2+i\epsilon\right]}
={-i\over (4\pi)^{n/2}}\left( M^2\right)^{n/2-1}\Gamma\left( 1-n/2\right)
\label{int01}
\ee

$$
J^{n+2}(11)={2\over (2\pi)^{n+1}}\int {d^{n+2}q\over \left[ q^2
+i\epsilon\right]\left[ (p+q)^2-M^2+i\epsilon\right]}=
{i\over (4\pi)^{n/2}}\left( M^2\right)^{n/2-1}\Biggl \{
{\Gamma\left( 1-{n/2}\right)\Gamma\left( n-1\right)\over 
\Gamma\left( n\right)}\times 
$$
\be
{ }_2F_1\left( 1,1-n/2;2-n; 1-z\right)+\Gamma\left( {n/2}\right)
\Gamma\left( 1-n\right)(1-z)^{n-1}
{ }_2F_1\left( {n/2}, n;n;1-z\right)\Biggr \}
\label{n2J11}
\ee

$$
J(11)={1\over (2\pi )^n}\int {d^nq\over \left[ q^2+i\epsilon
\right]\left[ \left( p+q\right)^2-M^2
+i\epsilon\right]}={i\left( M^2\right)^{n/2-2}\over (4\pi )^{n/2}}\Biggl \{
{\Gamma\left( 2-{n/2}\right)\Gamma\left( n-3\right)\over 
\Gamma\left( n-2\right)}\times 
$$
\be
{ }_2F_1\left( 1,2-n/2;4-n; 1-z\right)+\Gamma\left( {n/2}-1\right)
\Gamma\left( 3-n\right)(1-z)^{n-3}
{ }_2F_1\left( {n/2}-1, n-2;n-2;1-z\right)\Biggr \}
\label{J11}
\ee

$$
J(1101)={1\over (2\pi )^{2n}}\int {d^nq_1d^nq_2\over \left[ q_1^2+i\epsilon
\right]\left[ q_2^2+i\epsilon \right]\left[ \left( p+q_1+q_2\right)^2-M^2
+i\epsilon\right]}
$$
$$
={\left( M^2\right)^{n-3}\over (4\pi )^{n}}
\Biggl \{ {\Gamma\left( 3-n\right)\Gamma^2\left( n/2-1\right)
\Gamma\left( 2-n/2\right)\over \Gamma\left( n/2\right)}+
{\Gamma\left( 4-n\right)\Gamma^2\left( n/2-1\right)
\Gamma\left( 3-n/2\right)\over \Gamma\left( n/2+1\right)}z
$$
\be
+{1\over 12}\left[ 6-2\pi^2+15z+12{\rm Li}_2(1-z)+6\ln (1-z)\left( {(1-z)^2
\over z}+2(1-z)+2\ln z\right)\right]\Biggr \}
\label{aJ1101}
\ee

$$
J^{\mu}(1101)={1\over (2\pi )^{2n}}\int {d^nq_1d^nq_2 \ q_1^{\mu}\over 
\left[ q_1^2+i\epsilon
\right]\left[ q_2^2+i\epsilon \right]\left[ \left( p+q_1+q_2\right)^2-M^2
+i\epsilon\right]}
$$
$$
=-p^{\mu}{\left( M^2\right)^{n-3}\over (4\pi )^{n}}
\Biggl \{ {\Gamma\left( 3-n\right)\Gamma\left( n/2-1\right)
\Gamma\left( 2-n/2\right)\Gamma\left( n/2\right)\over \Gamma
\left( n/2+1\right)}
$$
$$
+
{\Gamma\left( 4-n\right)\Gamma\left( n/2-1\right)
\Gamma\left( 3-n/2\right)\Gamma\left( n/2\right)\over 
\Gamma\left( n/2+2\right)}z
+{1\over 24}\Biggl [ {61\over 3}-2\pi^2-{40\over 3}(1-z)
-2(1-z)^2-2(1-z)^3
$$
\be
-2{(1-z)^4\over z}+12{\rm Li}_2(1-z)+\ln (1-z)\left( 
12[1-z+\ln z]+{6(1-z)^2\over z}-{2(1-z)^3\over z^2}\right)\Biggr ]\Biggr \}
\label{aJ2102}
\ee

\be
J^{\mu\nu}(1101)={1\over (2\pi )^{2n}}\int {d^nq_1d^nq_2 \ q_1^{\mu} \ q_2^{\nu}
\over 
\left[ q_1^2+i\epsilon
\right]\left[ q_2^2+i\epsilon \right]\left[ \left( p+q_1+q_2\right)^2-M^2
+i\epsilon\right]}=C_1g^{\mu\nu}+C_2p^{\mu}p^{\nu}
\label{defc2}
\ee

$$
C_2={\left( M^2\right)^{n-3}\over (4\pi )^{n}}
\Biggl \{ {\Gamma\left( 3-n\right)
\Gamma\left( 2-n/2\right)\Gamma^2\left( n/2\right)\over \Gamma
\left( n/2+2\right)}
+
{\Gamma\left( 4-n\right)
\Gamma\left( 3-n/2\right)\Gamma^2\left( n/2\right)\over 
\Gamma\left( n/2+3\right)}z
$$
$$
+{1\over 72}\Biggl [ {241\over 12}-2\pi^2-{5\over 2}(1-z)
-{5\over 2}(1-z)^2-{3\over 2}(1-z)^3-{5\over 2}{(1-z)^4\over z}+
{(1-z)^4(1+z)\over z^2}
$$
\be
+12{\rm Li}_2(1-z)+\ln (1-z)\left( 
12[1-z+\ln z]+{6(1-z)^2\over z}-{2(1-z)^3\over z^2}+{(1-z)^4\over z^3}
\right)\Biggr ]\Biggr \}
\label{aC2}
\ee

\end{document}